\documentclass[preprint]{aastex}
\usepackage{graphicx}
\usepackage{url}
\begin{document}
\title
{Exo-Earth/Super-Earth Yield of JWST plus a Starshade External Occulter }
\author
{Joseph Catanzarite and Michael Shao}

\affil{ Jet Propulsion Laboratory, California Institute of Technology \break
4800 Oak Grove Drive, Pasadena, CA 91109, USA  }
\email{joseph.catanzarite@jpl.nasa.gov, michael.shao@jpl.nasa.gov}

\begin{abstract}

We estimate the exo-Earth/super-Earth yield of an imaging mission which combines the James Webb Space Telescope (JWST) with a starshade external occulter 
under a realistic set of astrophysical assumptions. For the purpose of this study, we define `exo-Earth' and `super-Earth' as a planet of mass 1 to 2  $M_{\oplus}$ and 2 to 10 $M_{\oplus}$, respectively, orbiting within the habitable zone (HZ) of a solar-type star. We show that for a survey strategy that relies on a single image as the basis for detection, roughly half of all exo-Earth/super-Earth detections will be false alarms, for $\eta_{\oplus}$ of 0.1, 0.2 and 0.3. Here, a false alarm is a mistaken identification of a planet as an exo-Earth/super-Earth, and we define $\eta_{\oplus}$ as the frequency of exo-Earth/super-Earths orbiting sun-like stars.

We then consider two different survey strategies designed to mitigate the false alarm problem. The first is to require that for each candidate exo-Earth/super-Earth, a sufficient number of detections are made to measure the orbit.  When the orbit is known we can determine if the planet is in the habitable zone. With this strategy, we find that the number of exo-Earth/super-Earths found is on average 0.9, 1.9 and 2.7 for $\eta_{\oplus} = 0.1, 0.2$ and $0.3$. There is a $\sim$40\% probability of finding zero exo-Earth/super-Earths for $\eta_{\oplus} = 0.1$.

A second strategy can be employed if a space-based astrometry mission capable of sub-microarcsecond precision has identified and measured the orbits and masses of the planets orbiting nearby stars. In this case, the occulter mission is much more efficient, because it surveys only the stars known to have exo-Earth/super-Earths.   We find that with prior knowledge from a space-based astrometric survey of 60 nearby stars, JWST plus an external occulter can obtain spectra, as well as orbital solutions, for the majority (70\% to 80\%) of the exo-Earth/super-Earths orbiting these 60 stars. The yield of exo-Earth/super-Earths is approximately five times higher than the yield for the JWST plus occulter mission without prior astrometric information. With prior space-based astrometry, the probability that an imaging mission will find zero exo-Earth/super-Earths is reduced to $<1\%$ for the case of $\eta_{\oplus} = 0.1$.

\end{abstract}

\section{Introduction}
Many instruments have been proposed to directly detect the reflected light from an Earth orbiting a nearby star. Nulling interferometers, and internal and external coronagraphs all have an inner working angle (IWA) that limits the region near a star within which a planet can be seen to the needed contrast level. The star-planet separation has to be larger than the IWA for the planet to be detected. When estimating how many exo-Earth/super-Earths such instruments would detect, previous direct imaging mission modeling studies commonly assume that each star has either zero planets or one exo-Earth/super-Earth planet \cite{brown05, brown10, sav08, sav09a, sav09b, sav10}. In some of these studies, it is also assumed that a single image of a planet is sufficient to confirm whether or not it is an exo-Earth/super-Earth.

Neither of these two assumptions is realistic, and they result in over-estimation of the science capability of an imaging mission. Recent RV (radial velocity) planet surveys of solar-type stars  reveal a significant population of Neptune and super-Earth planets in close orbits \cite{mayor09a, mayor09b}. Microlensing surveys, with more limited statistics, indicate that Neptune-mass planets within a few AU (cold Neptunes) are common \cite{gould10}. They are estimated to occur at a rate of 0.36 per $dex^{2}$ in the phase space of mass and semimajor axis (a dex is a decade interval on a logarithmic domain). This gives about 11\% probability of finding a Neptune-mass (10 to 100 $M_{\oplus}$) planet in the habitable zone. If planets in the terrestrial mass range (1 to 10 $M_{\oplus}$) follow the same distribution, one might expect that $\eta_{\oplus} \sim0.1$. An image of a Neptune that is 2 AU from the star could, depending on orbital inclination and orbital phase, have the same brightness and angular separation as an Earth in a 1 AU orbit. A detection from a single image can only be an exo-Earth/super-Earth candidate. It may be a false alarm, i.e. an exo-Earth/super-Earth candidate that is not an exo-Earth/super-Earth. We show that this ambiguity has enormous impact on the number of exo-Earth/super-Earths that could be found by direct detection missions, and needs to be carefully considered in the selection of a survey strategy.

In their recent paper, \citet{brown10} calculated that if $\eta_{\oplus} =0.3$,
that is, if 30\% of solar-type stars have an Earth in the habitable zone, then the combination of JWST and a starshade external occulter would on average find $\sim$5 Earths. The calculations were based on a set of well-defined assumptions, including the ones discussed above. We list their other main assumptions in Section~\ref{assumptions}. In this paper we introduce more realistic planet distributions into modeling of a JWST plus external occulter mission, and investigate the consequences. Based on the characteristics of the known population of exoplanets \cite{cum08}, we assume the distribution of planets is flat in both log(semi-major axis) and log(mass). For example, there are as many Earths between 0.7 and 1.4 AU as there are between 0.35 and 0.7 AU, and an equal number between 1.4 and 2.8 AU and so forth.  We also include the population of Neptune-mass planets that is referred to above. A discussion of our simple planet generating model is provided in an earlier paper \cite{shao10}.

The plan of this paper is as follows: the main assumptions about the JWST plus starshade external occulter \cite{brown10} are recapped in section~\ref{assumptions}.  Criteria for detection of an exo-Earth/super-Earth candidate in an image are discussed in section ~\ref{criteria}.  The false alarm problem in the detection of exo-Earth/super-Earth by imaging missions is illustrated in section ~\ref{misID}. Three possible observing strategies for external occulter missions, including two designed to mitigate false alarms, are outlined section~\ref{strategies}.

In section~\ref{detect-once} we analyze the results of a design reference mission similar to the one analyzed in \citet{brown10}, but with the more realistic distribution of planets, for three different values for $\eta_{\oplus}$: 0.1, 0.2 and 0.3.  For each case, we calculate the number of exo-Earth/super-Earth candidates detected and the probability that an exo-Earth/super-Earth candidate is a false alarm, i.e., not an exo-Earth/super-Earth.

In section~\ref{orbit} we investigate how effectively the JWST plus starshade external occulter can measure the orbits of the planets it detects.  A minimum of four detections are needed to determine an orbit \cite{shao10, glassman07}. After detecting the planet once, the instrument is required to detect the planet at three additional positions in its orbit, to confirm that the planet is in the habitable zone. We determine how many habitable zone orbits would be measured, for $\eta_{\oplus} = 0.1, 0.2$ and $0.3$. We also calculate the probability that no exo-Earth/super-Earth would be detected.

 How important is prior information on the content of nearby exoplanetary systems? Section~\ref{astrometry} deals with the scientific effectiveness of the JWST plus starshade external occulter if the stars with known exo-Earth/super-Earths were first identified by a space astrometry mission that measures their masses and orbits.  For the cases $\eta_{\oplus} = 0.1, 0.2$ and $0.3$, we determine the number of exo-Earth/super-Earth discoveries expected,  as well as the probability that no exo-Earth/super-Earths are detected.

\section{Assumptions for the JWST plus starshade external occulter mission}
\label{assumptions}

The main assumptions in \citet{brown10} were:
\begin{enumerate}
  \item IWA = 85 mas

  \item $\Delta mag \leq 26$, where  $\Delta mag$ is the magnitude difference between the planet and the star.  A planet can be detected if its flux contrast is brighter
	than $4\times10^{-11}$. Note that the Earth-Sun flux contrast is $1.2\times10^{-10}$ when Earth is at quadrature phase (i.e. half-illuminated).
  \item The total number of retargeting maneuvers, or distinct observations is $\sim$70 for the entire survey. This number is limited by the starshade propellant.
  \item The total number of stars in the target list is 26. This number is set by the requirement that a complete observation sequence, including imaging, instrument alignment, and spectra, may be done in the time during which the star is observable.

\end{enumerate}

Since the imaging observations are mostly in the JWST red (700 nm) and infrared (1150 nm) filters \cite{brown10}, we adopt a value of 0.2 for Earth's albedo. For gas-giant planets we adopt an albedo of 0.5, based on studies of solar system planets \cite{tra03}.

We did not look in detail at the Sun-exclusion issues. For JWST the angle between the target star and the Sun has to be between 85$\degr$ and 105$\degr$ \cite{brown10}. The first constraint is to keep sunlight out of the telescope and the second is to keep sunlight off the side of the starshade facing the telescope.
This is quite a bit more restrictive than other external occulter missions, which can look as close as 45$\degr$ to the Sun. When the Sun-exclusion issue is accounted for, the slew to the next target may take longer and use more propellant.

Without Sun-exclusion, for 26 stars spread uniformly across the sky, the next target is on average ~40$\degr$ away. With Sun-exclusion, the average distance will be somewhat larger. Sun-exclusion is more serious when we consider repeated observations of the much smaller number of stars with exo-Earth/super-Earth candidates to get the orbits of the planets. Without Sun-exclusion, for 8 to 10 stars uniformly spread over the sky, the average angle between targets is $\sim90\degr$. A starshade spacecraft that could accommodate $\sim$70 retargets with an average of 40$\degr$ between targets may only have $\sim$50 retargeting maneuvers if the average distance between targets was 90$\degr$. It would take the starshade $>3$ weeks on average \cite{glassman07} to move between targets. With the Sun-exclusion constraint, only 18\% of the sky is observable at any given time, so the issue of target availability will greatly increase the difficulty of follow-up measurements. In the present study we did not account for any of these limitations.

\section{Detection criteria for an exo-Earth/super-Earth candidate}\label{criteria}
We assume the contrast limit is $4\times10^{-11}$, and the IWA is 85 mas, as in \citet{brown10}. A planet can be detected if its angular separation from the star exceeds the IWA and its contrast exceeds the contrast limit. We take the inner and outer edges of the habitable zone to be $IHZ = 0.8\sqrt{L}$ AU and $OHZ = 1.6\sqrt{L}$ AU \cite{lun08}, where $L$ is the star's luminosity in solar units.

When a planet orbits a star, a coronagraph detects light from the star that is reflected by the planet. The flux contrast of the planet is determined by its distance from the star and the angle $\alpha$ from the star to the planet to the observer. If $i$ is the inclination of the planet's orbit, and $\phi$ is the orbital phase, then $cos(\alpha)= -sin(\phi)  sin(\it i)$. Assuming the planet is a Lambertian scatterer, and the orbit is circular, the planet-star flux contrast is given by \cite{brown05}
\begin{equation}
    F(\alpha)=\frac{2}{3}a\left(\frac{r}{R}\right)^{2} \frac{sin(\alpha)+(\pi-\alpha)cos(\alpha)}{\pi},
\end{equation}
where $R$ is the distance from the star to the planet, $r$ is the radius of the planet, and $a$ is the planet's albedo.

When $\alpha = \pi/2$, the planet appears half-illuminated (quadrature phase), and

\begin{equation}
    F=\frac{2a}{3\pi}\left(\frac{r}{R}\right)^{2}
\end{equation}

For each star, we can estimate the maximum flux contrast for a super-Earth at quadrature. The maximum flux $F_{max}$ is that of a rocky $10M_{\oplus}$ super-Earth at the IHZ. The radius of such a planet is $r\sim10^{1/3}r_{\oplus}=2.15r_{\oplus}$, where $r_{\oplus}$ is the radius of Earth. We estimated the radii of planets more massive than super-Earths from Figure 3 in \citet{cha09}.

The maximum flux is:

\begin{equation}
    F_{max}=\frac{2a}{3\pi}\left(\frac{r}{IHZ}\right)^{2}.
\end{equation}

If a detected planet's angular separation from the star is within the angular radius of the OHZ, then the planet could be in the habitable zone. If in addition, the flux is below $F_{max}$, then the planet could be an exo-Earth/super-Earth, and is identified as an exo-Earth/super-Earth candidate.

Therefore, the detection criteria for an exo-Earth/super-Earth candidate are:
\begin{enumerate}

\item $\theta < \frac{R_{OHZ}}{D}$, where $\theta$ is the star-planet angular separation in arcseconds, $D$ is the distance to the star in pc, and $R_{OHZ}$ is the radius of the outer habitable zone, in AU.
\item $F < F_{max}$, where $F$ is the flux contrast of the planet.
\end{enumerate}

\section{Confirmation of exo-Earth/super-Earth candidates}\label{misID}
If a planet is counted as an exo-Earth/super-Earth detection based on a single image in which its flux contrast and angular separation are consistent with those of a terrestrial planet in the habitable zone, there is a chance that it is a false alarm, i.e., that it is not a terrestrial planet and/or is not in the habitable zone. Without measuring its orbit, we cannot know whether or not it is in the habitable zone, and without measuring its mass, we cannot know whether or not it is a terrestrial planet. One visit is not enough.

\begin{figure}
\includegraphics{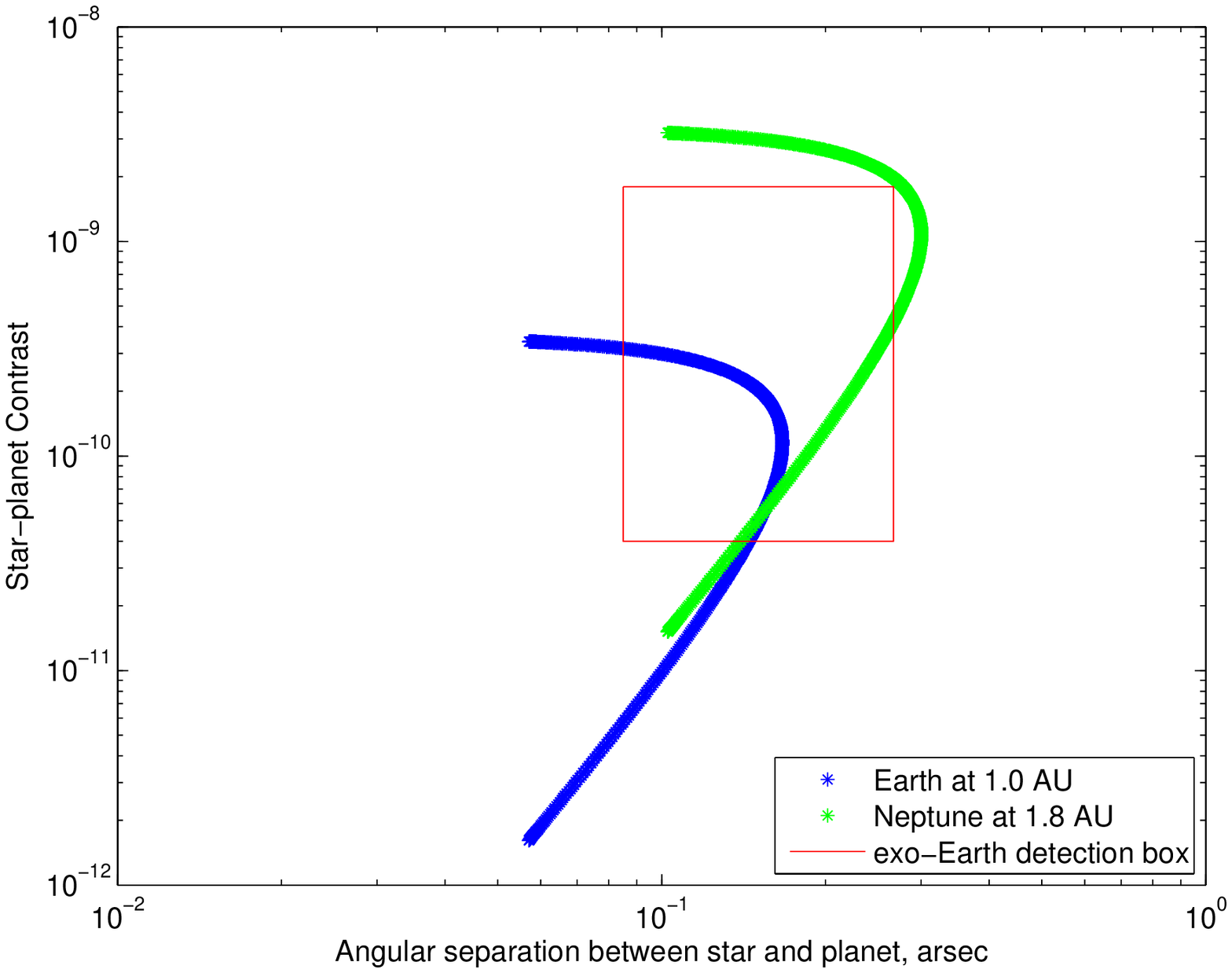}
\caption{Angular separation vs. contrast for Earth and a cold Neptune, at 6 pc.  The planets' trajectories overlap over a portion of their orbits. In fact, simulations show that most of the phase space inside the detection box could be occupied by either Neptune-mass planets at orbital semi-major axes ranging from 0.7 to 2.5 AU, or exo-Earth/super-Earths. Because Neptunes share the detectable phase space domain of exo-Earth/super-Earths, a Neptune-mass planet cannot be distinguished from an exo-Earth/super-Earth, given only one image. However, a prior astrometric mission could provide not only the orbital elements but also the mass of the planet, removing the ambiguity. The upper boundary of the detection box is the maximum flux contrast of a terrestrial planet ($10 M_{\oplus}$ at the inner edge of the habitable zone). The lower boundary is the minimum detectable flux contrast, set by the instrument. The left boundary is the IWA, and the right boundary is the outer edge of the habitable zone.}\label{fig:2planets}
\end{figure}

Figure~\ref{fig:2planets} shows the brightness versus angular separation of two planets that orbit a star at 6 pc. One is an Earth-mass planet at 1 AU, the other is a Neptune-mass planet at 1.8 AU; both orbit at an inclination of 70$\degr$. A portion of the trajectories of the two planets overlap. In fact, simulations show that most of the phase space inside the detection box could be occupied by either Neptune-mass planets at orbital semi-major axes ranging from 0.7 to 2.5 AU, or exo-Earth/super-Earths. Because Neptunes share the detectable phase space domain of exo-Earth/super-Earths, a Neptune-mass planet cannot be distinguished from an exo-Earth/super-Earth given only one image. If we were to take the spectrum of such a planet we might erroneously conclude that this is an exo-Earth/super-Earth at 1 AU with a methane atmosphere, rather than that this is a Neptune at 1.8 AU with a methane atmosphere. However, a prior astrometric mission could provide not only the orbital elements but also the mass of the planet, removing the ambiguity. Color is a diagnostic of the atmosphere of an exoplanet, but our knowledge of the colors of exoplanets is very rudimentary; for exoplanets in the HZ there is currently no data. The use of color to infer size or albedo contributes very little to the confirmation process.

On the other hand, prior RV data does not remove the ambiguity. Capability of RV for detection of exo-Earth/super-Earths depends on the three contributing factors to noise in RV measurements: photon noise, instrumental noise, and astrophysical noise. Photon noise can be `beaten down' by averaging many measurements together. Instrumental noise is reduced by efforts to control thermal and mechanical effects, wavelength calibration etc. Astrophysical noise sources include starspots, granulation, p-modes, and meridional flows.  Most of these effects have timescales that are long compared to RV measurements; as they cannot be averaged away with multiple measurements, they represent an ultimate limiting noise floor. Starspots on the Sun impose an RV jitter of 0.3 to 0.4 m/s \cite{mak09, cat07}, and the Sun's meridional flows contribute RV variability of 1.4 m/s RMS on on time scales of 0.1 to 1 year \cite{mak10}. Recent Kepler results \cite{cia10} indicate that the median photometric jitter for dwarf stars is 0.4 mmag, quite similar to that of the Sun. According to a simple scaling relation between photometric and RV jitter \cite{mak09}, the expected median RV jitter for dwarf stars is 0.8 m/s. Currently, HARPS \cite{may03} and HIRES \cite{Vogt94} are the best RV instruments.  Combined instrumental and astrophysical noise for HARPS and HIRES is 1 to 3.5 m/s, the RMS jitter remaining in the RV data after planets have been fitted out \cite{how10a}. The lowest RV signal that has been detected to date is 1.9 m/s, for GJ 581e \cite{mayor09a}, and for HD 156668b \cite{how10b}. ESPRESSO, the next-generation RV instrument, will come online at the VLT in 2015 and will be capable of RV precision below 10 cm/sec.   Though this instrument would be capable of detecting the Earth orbiting the Sun in the absence of astrophysical noise, Kepler data shows that sun-like stars with astrophysical noise significantly below 1 m/s are not common.

An Earth at 1 AU has a RV signal of $\sim$0.1 m/s, which is likely to be undetectable by RV. A Neptune-mass planet ($17M_{\oplus}$) at 1.8 AU has a RV signal of $\sim$$\frac{17}{\sqrt{1.8}}\times0.1$ m/s $\sim$1.3 m/s, which is just below the currently detectable range. Even if a prior RV survey had discovered a Neptune orbiting the star at 1.8 AU, given a single image showing a candidate exo-Earth/super-Earth, one could not distinguish whether it was a bona fide exo-Earth/super-Earth or the Neptune. Astrometry provides a direct measurement of mass, which is not only a critical science parameter, but is key to the confirmation of an exo-Earth/super-Earth from direct imaging observations.

\section{Observing strategies for external occulter missions}
\label{strategies}

\citet{brown10} described an optimized observing strategy to observe target stars until the planet was detected once, outside of the IWA, and then to take the spectrum immediately after the detection.  In Section ~\ref{detect-once}, we employ a similar strategy with the JWST plus starshade external occulter and use Monte Carlo simulations to calculate the probability that a planet that was identified as an exo-Earth/super-Earth candidate either wasn't in the habitable zone, or was a cold Neptune instead of a terrestrial planet. We shall refer to this strategy as the \emph{detect-once survey}.

\citet{kasdin10} described a two part observing strategy, with a search phase to detect any planet once, followed by a series of repeated observations to measure the orbit. We shall refer to this strategy as the \emph{orbit measurement survey}. In Section~\ref{orbit}, we apply a similar strategy to the JWST plus starshade external occulter. Given a survey mission with a maximum of 70 retargeting maneuvers, we use Monte Carlo simulations to  determine how many exo-Earth/super-Earths would have their orbits measured. We calculate the probability that zero exo-Earth/super-Earths are found and confirmed to have habitable zone orbits.

A more promising observing strategy is to detect the exo-Earth/super-Earths with a space astrometry mission prior to the direct imaging survey with the JWST plus starshade external occulter.  Knowing beforehand where the exo-Earth/super-Earths are gives two advantages which save observing time and therefore increase the yield of exo-Earth/super-Earths. First, the imaging/spectroscopy survey need only include stars known to have exo-Earth/super-Earths. Second, we need only image the planet twice (rather than the four times needed in an \emph{orbit measurement survey}) in order to confirm it. We shall refer to this strategy as the \emph{orbit confirmation survey}. We apply this strategy to the JWST plus starshade external occulter in Section~\ref{astrometry}.

Each of the observing strategies are described in more detail below, followed by a discussion of the associated Monte Carlo simulation results.

\section{The \emph{detect-once} survey with a JWST plus starshade external occulter mission}
\label{detect-once}

In the observing scenario described in \citet{brown10}, the 26 targets are observed on a schedule designed to maximize the number of planet discoveries, where a single image of a planet is counted as a discovery. If a planet is detected, a spectrum is taken, and that star is removed from the list. Their adaptive scheduling algorithm is used to determine when and if each star should be revisited. They applied their algorithm only for the case of $\eta_{\oplus} = 0.3$, for which they found 5.1 exo-Earth/super-Earth discoveries.  The JWST observations are mostly in the 700 nm and 1150 nm filters, for which the Earth's albedo is close to 0.2.

We repeated their study, first under their assumption that $\eta_{\oplus}= 0.3$ and that each star has either zero planets or one exo-Earth/super-Earth planet and then under the assumption of the more realistic planet distribution described in our previous paper \cite{shao10}. The simulations were run for cases of $\eta_{\oplus}$ of 0.1, 0.2 and 0.3.

We used the same 26 stars in our simulation of the JWST mission. For our study, we did not fully optimize the observing strategy. Instead we cycled through the target list, ordered by the likelihood to detect on the next visit an Earth-mass planet at the habitable zone scaled by $\sqrt{L}$.  The likelihood of detection on the next visit is estimated by assuming that
\begin{itemize}
  \item The star has a planet at the scaled 1 AU habitable zone with a probability of $C$ to see the planet on a randomly-timed visit.
  \item Consecutive visits are independent.
\end{itemize}

If $C$ is the single-visit completeness, the fraction of random visits to the star that would result in a detection of an Earth-mass planet at the scaled habitable zone \cite{brown05}, then the probability that the planet will not be detected on a single random visit is $1-C$; and the probability that it will not be detected on N consecutive random visits is $(1-C)^{N}$. If the planet has not been detected on N consecutive visits, the probability to detect it on the next visit is estimated as  $P_{next} = C(1-C)^{N}$. If the star has had N previous visits with no detection, the probability for detection on the next visit is estimated as $P_{next} = C(1-C)^{N}$.
We have neglected the possibility of correlation between visits. If the planet is not detected on the first visit, then if a followup visit is made very soon it is likely to be missed again, since it hasn't moved much. \cite{brown10} Figure 1, shows that if the delay is long enough (roughly 2 months between visits), the subsequent visit becomes (to a good approximation) uncorrelated from the first.

To estimate $C$ for an exo-Earth/super-Earth at the HZ of a given star one first generates an ensemble of orbits at inclinations that are randomized over cosine(i). For each orbit, determine the fraction of time $f$ that the exo-Earth/super-Earth is both outside the IWA and has flux contrast exceeding the minimum detectable contrast. Average $f$ over the inclinations to estimate $C$.  In this study, we estimate $C$ by setting the inclination of the planet's orbit to its most likely value of $60\degr$. We find good agreement with the values determined in \cite{brown10}.

If at a given star, an exo-Earth/super-Earth candidate is detected, based on its flux, assumed albedo, and angular separation, that star is moved to the candidate list and is not observed again. At the completion of each cycle, the list is rank-ordered by the probability of detection for the next cycle of visits. We advance through the list in this way until the number of visits reaches 70.

We ran Monte Carlo simulations consisting of 1000 survey realizations each under the assumption of realistic planet distributions, for cases of $\eta_{\oplus}$ = 0.1, 0.2 and 0.3. Our results are in Table~\ref{table-detection}. For comparison with the simulations in \citet{brown10}, we also show the results of our simulations under the assumption that each star has either zero planets or one exo-Earth/super-Earth planet.

For $\eta_{\oplus} = 0.3$ and only exo-Earth/super-Earth planets in the simulation, our result was 5.7 exo-Earth/super-Earth discoveries, which is roughly consistent with the result of 5.1 exo-Earth/super-Earths from \citet{brown10}; our result is higher because we did not account for solar exclusion constraints in the observing cadence.

The \emph{detect-once survey} strategy is problematic because of the lack of confidence that an exo-Earth/super-Earth candidate detected in a single image is really an exo-Earth/super-Earth. For $\eta_{\oplus} = 0.1, 0.2$ and 0.3, we find probabilities of 58\%, 52\% and 49\%, respectively that the exo-Earth/super-Earth candidate is a false alarm, i.e., not an exo-Earth/super-Earth.  For the case of $\eta_{\oplus} = 0.1$, there is a $\sim$14\% chance that the detected exo-Earth/super-Earth candidates include no exo-Earth/super-Earths.

\section{The \emph{orbit measurement survey} with a JWST plus starshade external occulter mission}\label{orbit}

For this survey, we followed an observing strategy designed to measure the orbits of exo-Earth/super-Earth candidates, similar to the one outlined in the paper on the proposed Orbiting Ozone Observatory \cite{kasdin10}. We again used the target list of 26 stars \citet{brown10}.

The observing strategy has two parts. The first part is an initial survey to observe all 26 stars once and identify those with candidate exo-Earth/super-Earths. The second part is to follow up on the stars with candidate exo-Earth/super-Earths with the goal of obtaining enough observations to measure their orbits.

Nine stars are designated as follow-up stars. Selected first for the follow-up list are stars at which a candidate exo-Earth/super-Earth was found at the first visit.  If the initial survey turned up fewer exo-Earth/super-Earth candidates than the desired number of follow-up stars, the follow-up list is filled out with stars at which no exo-Earth/super-Earth candidate was found, selected by rank order of the probability of detecting an exo-Earth/super-Earth at the next visit. As in Section~\ref{detect-once}, if there is no detection at the first visit to a star, the probability for detection on the next visit is estimated as $P_{next} = C(1-C)$.

The total number of observations in the survey is ultimately limited to $\sim$70 by the starshade propellant supply \cite{brown10, glassman07}. After the initial cycle of 26 observations, $\sim$44 additional visits are allowed; so the number of followup stars times the number of additional observations of each followup star equals $\sim$44. We found that 9 followup stars with 5 additional visits to each was near-optimal. Each of the 9 stars is observed a total of 6 times.

The observing strategy could be improved by ceasing to follow up on a star when the exo-Earth/super-Earth candidate has already been detected four times, or when it becomes clear that it cannot be detected four times.
If an exo-Earth/super-Earth candidate is detected on the first visit, then

\begin{enumerate}
  \item If it is either detected on each of the first 3 followup visits or missed on each of the first 3 followup visits, then there is no need to observe it again, so we could save the last 2 visits.

  \item If the first 3 followup visits are not all hits and not all misses, then: if the exo-Earth/super-Earth is either detected in 3 of the first 4 followup visits, or missed in 3 of the first 4 followup visits, there is again no need to observe it again, so we could save the last visit.
\end{enumerate}

This allows an extra visit for some of the 9 followup stars, which would increase the exo-Earth/super-Earth yield, but not significantly enough to change the major conclusions of this study. We did not implement this improvement, because the current observing strategy does not impose the severe solar exclusion constraints on target availability, thereby already over-estimating the exo-Earth/super-Earth yield.

We ran Monte Carlo simulations consisting of 1000 survey realizations each under the assumption of realistic planet distributions, for cases of $\eta_{\oplus}$ = 0.1, 0.2 and 0.3. The 26 stars are each observed once in an initial survey. Next, 9 followup stars are selected and observed an additional 5 times each, so that the total number of visits to each is 6.

Results are shown in Table~\ref{table-meas_conf}. For every exo-Earth/super-Earth detection, the measured orbit is confirmed to lie within the HZ, and, although no mass measurement is possible, the measured flux is consistent with that of a terrestrial planet. Therefore there are no false alarms in the \emph{orbit measurement survey}.

\begin{figure}
\includegraphics{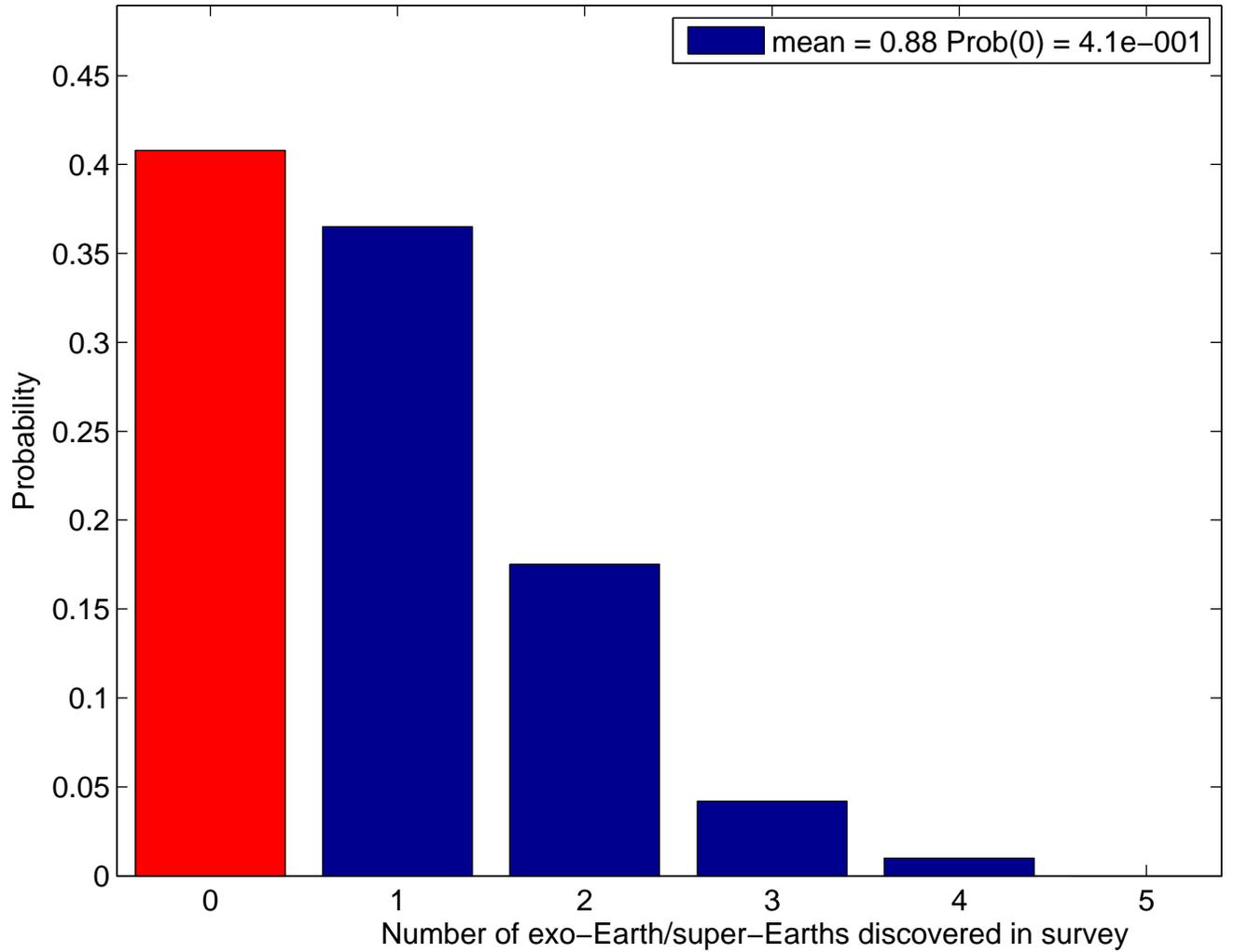}
\caption{Histogram of detected exo-Earth/super-Earths for the \emph{orbit measurement survey} of 26 stars. For $\eta_{\oplus} = 0.1$, there is a 41\% chance that the \emph{detect-once survey} will find no exo-Earth/super-Earths. }\label{fig:orbit-meas}
\end{figure}

The \emph{orbit measurement survey} is risky for low $\eta_{\oplus}$, because the expected number of exo-Earth/super-Earth detections is small, and the probability of failing to detect an exo-Earth/super-Earth is significant. Figure~\ref{fig:orbit-meas} shows that for the case of $\eta_{\oplus}$ of 0.1, we can expect to find 0.9 exo-Earth/super-Earths, and there is a 41\% probability to find no exo-Earth/super-Earths.

\section{The \emph{orbit confirmation survey} with a JWST plus starshade external occulter mission}\label{astrometry}

A space-based astrometry mission capable of submicroarcsecond precision can detect $\sim95$\% of the exo-Earth/super-Earths orbiting 60 nearby stars and measure their orbits and masses \cite{tra10, unwin08}.  With this information in hand, an external occulter mission to detect exo-Earth/super-Earths benefits in two ways. First, the instrument need only spend its time at stars at which exo-Earth/super-Earths are known to exist. This is a smaller and richer list; for $\eta_{\oplus}$ of 0.3, the list would contain 18 stars with exo-Earth/super-Earths on average. By comparison, the list of 26 stars would only contain about 8 stars with exo-Earth/super-Earths, on average. Secondly, only two detections of an exo-Earth/super-Earth are sufficient: the first to refine the orbital phase, and the second to confirm the orbital solution. For a discussion, see \citet{shao10}. The imaging mission alone provides no information on the mass of the planet; the prior astrometry mission measures the mass of the exo-Earth/super-Earth candidate, allowing us to know whether it is indeed an exo-Earth/super-Earth.

For the target list for this survey, we started with a database of 120 nearby FGK target stars for external occulter missions that was adopted for the Orbiting Occulter Observatory (O3), \cite{kasdin10}, kindly provided by Doug Lisman. To make a list for the \emph{orbit confirmation survey}, we sorted these by single-visit completeness (Brown 2005), for detection of an Earth-mass planet in the HZ by imaging with an IWA of 85 mas. We then selected the best 60 stars for which an Earth-mass planet could be detected by space-based astrometry at SNR of 5.8 at the inner edge of the habitable zone. The most distant star in the list of 26 JWST targets proposed in the \citet{brown10} is HIP 86614, at a distance of 22 pc. All of the stars in the best 60 targets for a space-based astrometry mission are closer than 22 pc; therefore we assume each can be observed by JWST in the time frame allotted in Table 2 from \citet{brown10}.

We assume that a space-based astrometry mission has previously surveyed these stars and has measured orbits and masses of 95\% of the exo-Earth/super-Earths. Typical errors in the fitted orbit parameters are found in \cite{tra10b}. For detections with SNR above 5.8, median period and mass uncertainties were 1\% and 3\% respectively. For planets close to the detection limit, mass uncertainty was generally better than about 30\%, and period uncertainty was better than about 4\%. In our simulated coronagraph search we use only the knowledge that an exo-Earth/super-Earth has been detected or not detected at a given star. We do not make use of the fitted parameters, so the errors on the fitted parameters do not propagate into the findings of the paper. If we had used prior knowledge of the fitted period to time the successive visits after the first detection, we'd expect some improvement in our results.

For $\eta_{\oplus}$ = 0.1, 0.2, and 0.3, we expect on average 6, 12, and 18 of these stars to have exo-Earth/super-Earths. The observing scenario is to cycle through the list of stars known to have exo-Earth/super-Earths, repeatedly observing each star until either an exo-Earth/super-Earth candidate is detected twice, or a maximum number of observations is reached without getting two detections of an exo-Earth/super-Earth candidate.  The optimum value for the maximum number of observations per star depends somewhat on $\eta_{\oplus}$. Though it also may vary with the angular size of the HZ, with a large HZ requiring fewer observations, we used the same maximum number of observations for each star. We found that a maximum of 10 observations is close to optimal for $\eta_{\oplus}$  of 0.1, 0.2 and 0.3.

We ran Monte Carlo simulations consisting of 1000 survey realizations each, for  $\eta_{\oplus}$ of 0.1, 0.2 and 0.3. On average, the \emph{orbit confirmation survey} yields 4.7, 9.1 and 12.0 exo-Earth/super-Earths, respectively. On average, the fraction of exo-Earth/super-Earths recovered by the imaging survey is 80\%, 77\%, and 68\%, respectively. The decrease in the exo-Earth/super-Earth recovery fraction with increase in $\eta_{\oplus}$ is due to the effect of confusion. In our planet population model, the occurrence frequency of planets of all masses is correlated with that of exo-Earth/super-Earths. Increasing $\eta_{\oplus}$ therefore increases the likelihood that any planetary system with an exo-Earth/super-Earth also has a cold Neptune, which could be confused with the exo-Earth/super-Earth. The probability of finding zero exo-Earth/super-Earths is less than 1\% for $\eta_{\oplus}$ = 0.1, and less than 0.1\% for $\eta_{\oplus}$ = 0.2 and 0.3. Results are in Table~\ref{table-meas_conf}, which shows a side-by-side comparison with results of the \emph{orbit measurement survey}.

\section{Summary and conclusions}

Is it sufficient to measure the orbit of the exo-Earth/super-Earth and take its spectrum, without measuring its mass? Characterization of a potential exo-Earth/super-Earth should ideally include measurement of its mass, semimajor axis, orbital inclination and eccentricity, as well as information about the spectrum of its atmosphere.

We have evaluated three exo-Earth/super-Earth characterization strategies for a notional JWST plus external occulter exo-Earth/super-Earth characterization mission. The main constraint for an external occulter mission is the limitation of the total number of visits to stars, set by the supply of propellant needed to move and orient the starshade as it slews from one target to the next. We did not consider viewing constraints imposed by solar exclusion.
\begin{enumerate}
  \item A \emph{detect-once survey}, in which a single image of a planet is followed up immediately by a spectrum. This strategy was described by \citet{brown10}.
  \item An \emph{orbit measurement survey}, in which four images of a planet, sufficient for the measurement of its orbit, are obtained. The fourth image is followed by a spectrum. This strategy was described in \citet{kasdin10}.
  \item An \emph{orbit confirmation survey}, in which the exo-Earth/super-Earths are known from prior astrometry. In this case, only two images are needed: the first detection refines the orbital phase, and the second provides confirmation. The second image is followed by a spectrum.
\end{enumerate}

We found that the \emph{detect-once survey}, which relies on a single detection of an exo-planet, is inadequate, since there is a high likelihood that exo-Earth/super-Earth candidates will be  misidentified. About half or more (58\%, 52\% 49\%) of the exo-Earth/super-Earth candidates found are not in fact exo-Earth/super-Earths; they are either not in the habitable zone, outside the terrestrial mass range, or both. If $\eta_{\oplus} = 0.1$ there is a 14\% chance that there are zero exo-Earth/super-Earths among the exo-Earth/super-Earth candidates found.

The exo-Earth/super-Earths found in the \emph{orbit measurement survey} are known to be in the habitable zone, and to have flux contrast consistent with terrestrial planets, therefore they are not likely to be false alarms, as in the \emph{detect-once survey}. But such a survey is very inefficient, with yield of 0.9, 1.9, and 2.7 exo-Earth/super-Earths for $\eta_{\oplus}$ of 0.1, 0.2, and 0.3 respectively. The probability of finding zero exo-Earth/super-Earths is 41\% if $\eta_{\oplus}$ is 0.1.

Of the three survey strategies considered, the only one that can provide all the information needed to characterize an exo-Earth/super-Earth (including planet mass) is the \emph{orbit confirmation survey}. With prior knowledge of the planet populations of the 60 best imaging targets from a space-based astrometry mission, we find that an \emph{orbit confirmation survey} yields spectra, as well as mass and orbital information for the majority of the exo-Earth/super-Earths. The exo-Earth/super-Earth yield was 4.7, 9.1, and 12.0 exo-Earth/super-Earths for $\eta_{\oplus}$ of 0.1, 0.2, and 0.3 respectively. The exo-Earth/super-Earth yield is $\sim$5 times higher than that of the \emph{orbit measurement survey}. The probability of finding zero exo-Earth/super-Earths for the case of $\eta_{\oplus} = 0.1$ is less than 1\%.
Science ``risk" is a significant consideration. An occulter mission which is limited to a relatively small total number of visits ($\sim$100) is ultimately more suited to exo-Earth/super-Earth characterization than initial detection. The combination of astrometric and direct imaging instruments is the best way to maximize the science return and reduce the risk that the direct imaging mission fails to yield the spectrum of an exo-Earth/super-Earth.

\acknowledgments
The research described in this paper was carried out at the Jet Propulsion Laboratory, California Institute of Technology, under contract with the National Aeronautics and Space Administration. Copyright 2010 California Institute of Technology. Government sponsorship acknowledged.  We thank Doug Lisman for providing the target list for the Occulting Ozone Observatory mission. We thank Steve Edberg, Varoujan Gorjian, Jim Marr, Xiaopei Pan, and especially John Davidson and Steve Unwin for critical reading and constructive comments and suggestions. We wish to thank the referee for a valuable review that contributed to many improvements in the paper.

\clearpage
\begin{table}
\begin{center}
\caption{Mission modeling simulations for the JWST plus starshade external occulter: \emph{detect-once survey} of 26 stars. Roughly half of the detected exo-Earth/super-Earth candidates turn out to be false alarms, i.e. not exo-Earth/super-Earths. For $\eta_{\oplus} = 0.1$, there is a 14\% probability that no exo-Earth/super-Earths will be detected. Two cases were considered: a simplified case of an isolated exo-Earth/super-Earth either present or not present; and a realistic distribution of planets in orbit radius and mass.\label{table-detection}}
\vspace{0.07in}
\begin{tabular}{ccccccc}
\tableline\tableline
Realistic  & $\eta_{\oplus}$ & Number of                  & Number of & Number of & False alarm&Probability of\\
planet                              &      &   exo-Earth/ &exo-Earth/     & false alarms & probability &  no exo-Earth/         \\
distribution& &super-Earth  & super-Earths & & & super-Earth\\
& & candidates &  & & & detections\\
\tableline
N & 0.1 & 1.9 &  1.9 &  N/A &N/A &0.14  \\
\tableline
N & 0.2 & 3.8  &  3.8 &N/A & N/A &0.04 \\
\tableline
N & 0.3 & 5.7 &  5.7 &N/A & N/A &0.01  \\
\tableline
Y & 0.1 & 4.5 &  1.9&2.6& 0.58 &0.14  \\
\tableline
Y & 0.2 & 7.9  & 3.8 &4.1& 0.52 &0.02 \\
\tableline
Y & 0.3 & 11.0 & 5.6 &5.4 & 0.49 &0.001 \\
\tableline
\end{tabular}
\end{center}
\end{table}

\begin{table}
\begin{center}
\caption{Mission modeling simulations for the JWST plus starshade external occulter:  (a) \emph{orbit measurement survey} of 26 stars without prior astrometry. (b) \emph{orbit confirmation survey} of 60 stars, with prior knowledge from space-based astrometry.\label{table-meas_conf}}
\vspace{0.07in}
\begin{tabular}{ccccc}
\tableline\tableline
$\eta_{\oplus}$  & exo-Earth/ & exo-Earth/& Probability of no & Probability of no\\
& super-Earth& super-Earth&  exo-Earth/& exo-Earth/\\
& yield & yield&super-Earth& super-Earth\\
& & & detections&detections\\
&  (a) \emph{orbit measurement}               &(b) \emph{orbit confirmation}  & (a) \emph{orbit measurement}  &  (b) \emph{orbit confirmation}\\
& \emph{survey} &\emph{survey}& \emph{survey}  & \emph{survey} \\
\tableline
0.1 & 0.9 & 4.7  & 0.41& 0.007\\
\tableline
0.2 & 1.9 & 9.1  & 0.16& $<$0.001\\
\tableline
0.3 & 2.7 & 12.0  & 0.07 & $<$0.001\\
\tableline
\end{tabular}
\end{center}
\end{table}

\end{document}